\documentclass[twocolumn,showpacs,preprintnumbers,amsmath,amssymb]{revtex4-2}

\usepackage{graphicx}
\usepackage{color,xcolor}
\usepackage{lipsum}
\usepackage{tabularx}
\usepackage{array}
\usepackage{mathrsfs}
\usepackage{amssymb}
\usepackage{amsmath}
\usepackage{cases}  
\usepackage{graphicx}
\usepackage{subfigure} 
\usepackage{dcolumn}
\usepackage{bm}
\usepackage{verbatim} 
\usepackage[pdfstartview=FitH,colorlinks,linkcolor=blue,anchorcolor=blue,citecolor=blue]{hyperref}
\usepackage{amsthm}  
\usepackage{multirow} 
\usepackage{booktabs} 
\usepackage[section]{placeins}

\usepackage{physics}
\usepackage{soul}
\usepackage[marginal]{footmisc}

\begin{document}
	
\title{Quantum key distribution over scattering channel}



\author{Qi-Hang Lu${}^{1,2}$}
\author{Fang-Xiang Wang${}^{1,2,*}$}
\author{Kun Huang${}^{3}$}
\author{Xin Wu${}^{1,2}$}
\author{Shuang Wang${}^{1,2}$}
\author{De-Yong He${}^{1,2}$}
\author{Zhen-Qiang Yin${}^{1,2}$}
\author{Guang-Can Guo${}^{1,2}$}
\author{Wei Chen${}^{1,2,\dagger}$}
\author{Zheng-Fu Han${}^{1,2,\S}$}
 
\affiliation{${}^1$CAS Key Laboratory of Quantum Information, University of Science and Technology of China, Hefei 230026, China\\
	${}^2$CAS Center For Excellence in Quantum Information and Quantum Physics, University of Science and Technology of China, Hefei, Anhui 230026, China\\
        ${}^3$Department of Optics and Optical Engineering, University of Science and Technology of China, Hefei 230026, China\\
	${}^*$Corresponding author: fxwung@ustc.edu.cn\\
	${}^\dagger$Corresponding author: weich@ustc.edu.cn\\
	${}^\S$Corresponding author: zfhan@ustc.edu.cn
}


\date{\today}

\begin{abstract}
Scattering of light by cloud, haze, and fog decreases the transmission efficiency of communication channels in quantum key distribution (QKD), reduces the system's practical security, and thus constrains the deployment of free-space QKD. Here, we employ the wavefront shaping technology to compensate distorted optical signals in high-loss scattering quantum channels and fulfill a polarization-encoded BB84 QKD experiment. With this quantum channel compensation technology, we achieve a typical enhancement of about 250 in transmission efficiency and improve the secure key rate from 0 to $1.85\times10^{-6}$ per sifted key. The method and its first time validation show the great potential to expand the territory of QKD systems from lossless channels to highly scattered ones and therefore enhances the deployment ability of global quantum communication network.
\end{abstract}

\pacs{Valid PACS appear here}
\maketitle


\section{\label{sec:level1}Introduction}
Quantum key distribution (QKD) \cite{bennett1984proceedings,xu2020secure} paves the way for two legitimate parties (conventionally called Alice and Bob) to share secret keys. For decades, QKD has developed rapidly from proof-of-principle experiments to commercial applications \cite{bennett1990experimental,peev2009secoqc,sasaki2011field,wang2014field,boaron2018secure,yuan201810,jacobs1996quantum,schmitt2007experimental,bedington2017progress,liao2017long,yin2020entanglement,cao2020long,chen2021integrated}. QKD is mainly implemented over fiber-based \cite{sasaki2011field,wang2014field,boaron2018secure,yuan201810} and free-space channels, which is the essential part of establishing a global secure quantum network. The free space QKD channel easily suffers from the influence of atmospheric turbulence and optical noise\cite{liao2017long,cao2020long,moschandreou2021experimental,erven2012studying,liorni2019satellite,pirandola2021limits}. Different from classical optical communication systems, the influence of a quantum channels cannot be compensated by enhancing the transmitted photon number, which decreases communication security. Previous studies used the adaptive tracking, atmospheric phase correction or post-processing methods to mitigate the influence of atmospheric turbulence and optical noise to QKD systems \cite{liao2017long,cao2020long,moschandreou2021experimental,erven2012studying,liorni2019satellite,pirandola2021limits} and have achieved significant progress during the last twenty years and has been successfully demonstrated between satellite and ground over a channel length from 10 km to 1000 km. \cite{jacobs1996quantum,schmitt2007experimental,bedington2017progress,liao2017long,yin2020entanglement,cao2020long,chen2021integrated,moschandreou2021experimental}. 

However, most free-space QKD systems were implemented under the conditions that the channel are clearly seen without haze or fogs and the channel loss is relatively low and mostly between 10 to 40dB \cite{schmitt2007experimental,liao2017long,cao2020long,chen2021integrated,moschandreou2021experimental}. In more complicated practical channel conditions with, such as, cloud, dust, haze or fogs, strong scattering effects may exist and leads to the large transmission loss \cite{deepak1982significance,colvero2005real,muhammad2007characterization,ijaz2013modeling,vasylyev2017free,grabner2013multiple}, which can be 60 dB or larger. Distributing secure key through the scattering-induced high-loss quantum channel remains to be verified for free-space QKD systems. In the strong scattering channel, the beam deformation is much stronger and the beam may even be destructed into speckles, making the performance of QKD system decline and even be difficult to share secret keys \cite{deepak1982significance,vasylyev2017free,grabner2013multiple,katz2012looking,zhao2020performance} . Therefore, an effective method for compensating the strong scattering effects and improving the channel performance is vital to free-space QKD systems.

For classical systems, wavefront shaping methods have been developed to deal with light field propagation through scattering media. Wavefront shaping can modulate the optical field by phase and (or) amplitude and hence compensate the scattering effect. Commonly used wavefront shaping methods include transmission matrix (TM) measurement \cite{popoff2010measuring,gong2019optical}, digital optical phase conjugation (DOPC) \cite{yaqoob2008optical}, and iterative algorithms \cite{vellekoop2007focusing,conkey2012genetic}. Hao et al. proposed a DOPC scheme to increase the channel efficiency of a classical optical communication system in 2014 \cite{hao2014self}. And the task of shaping or monitoring quantum states after scattering media has attracted much attention recently \cite{defienne2014nonclassical,lib2020real}.  However, applying the wavefront shaping methods in QKD through scattering channels has rarely been studied.

Compared to the TM and the DOPC method, the genetic algorithm (GA) developed by John Holland \cite{holland1992adaptation} has shown its advantage due to its global searching ability independent of the initial guess of the solution. Therefore, it has been applied to various of applications \cite{conkey2012genetic,deb2002fast,jones1997development,chiang2005improved,mahlab1991genetic}, e.g., shaping the focused spots through the scattering medium \cite{conkey2012genetic}. We implement QKD experiments through strong scattering quantum channels by utilizing the GA wavefront-shaping compensation approach. We applied the global searching and anti-noise merits of GA by monitoring the single-photon detecting signals of a QKD system, which means there is no need to increase the system complexity. Experimental results show that the channel transmission efficiency can be enhanced by about 250 times with the compensation. As a result, a secure key rate of $1.85\times10^{-6}$ per sifted signal is achieved while there is no secure key rate before optimization. Our work experimentally verifies the feasibility of distributing secret quantum keys through a strong scattering channel and will expand its application scope. 

\section{\label{sec:level1}Genetic Algorithm and Demonstration Systems}
\subsection{\label{sec:level2}Scheme of Algorithm and QKD system}
\begin{figure}[htb]
	\centering
	\includegraphics[width=9cm]{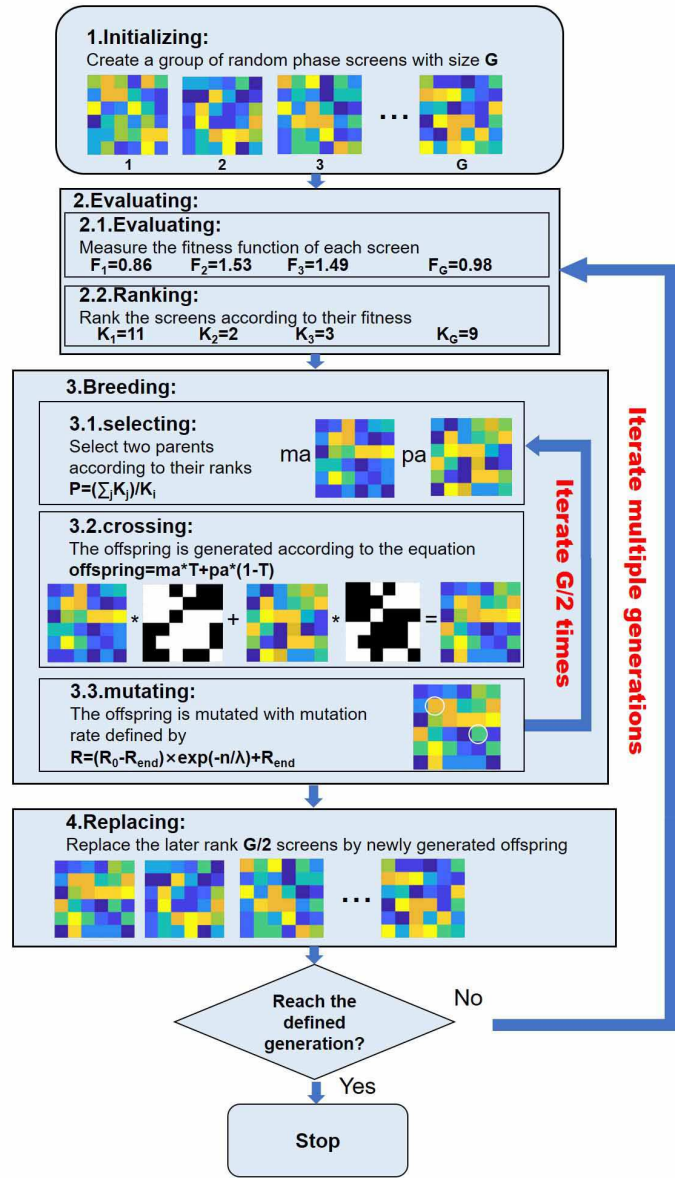}
	\caption{Flowchart of the GA-based wavefront-shaping compensation method.}
	\label{fig:1GA}
\end{figure}

Transmission efficiency over scattering channels should be improved while keeping a relatively low QBER to enhance the performance of the QKD. Thus, the compensation module is required to gather the transmitted photons inside a predefined area and keep the spot shape as close to Gaussian distribution as possible. GA searches for a better solution by emulating the evolution process in nature between adjacent generations. Compared to other wavefront methods, such as the TM and DOPC method, GA gives a global optimum without the initial guess of the scattering characteristic. Thus, GA is a valuable tool to compensate the channel effect and to optimize the optical field through a scattering media \cite{conkey2012genetic,vellekoop2015feedback}. As is shown in Fig. \ref{fig:1GA}, the primary process of this wavefront-shaping GA can be divided into four steps:

1. Initialization. At the beginning of optimization, a group of random patterns that offers the random initial search of the solution space is generated.

2. Evaluation and ranking. All the patterns are evaluated by the predefined fitness function. Then the patterns are ranked according to evaluation value, with a better fitness of a minor rank (ranking). 

3. Breeding. The process of generating offspring patterns from initial patterns (the so-called parent patterns) is then executed. Firstly the parents are selected by using a roulette method, which randomly selects two among all parent patterns in the group with the selecting probability according to the equation below:
\begin{equation}
P_i=\frac{\sum_j K_j}{K_i},
\end{equation}
where $K_i$ is the rank of the $i$-th pattern. At each time, one offspring pattern is generated according to the equation $offspring=T\times ma+(1-T)\times pa$, where $T$ is a random binary template and $ma$ and $pa$ are the parent patterns selected. The generated offspring, which combines the possible excellent parts of the parents, is then mutated by randomly changing the values of a few pixels of the pattern to expand the searching ability of the algorithm. In order to compromise between optimization speed and global searching ability, the mutation rate $R$ is set to be exponentially decaying, following $R=(R_0-R_{end})\times \exp(-n/\lambda)+R_{end}$, where $R_0$, $R_{end}$, and $\lambda$ are the initial mutation rate, the final mutation rate, and the decay factor, respectively.

4. Replacement. The parent patterns of the last half ranks are replaced by the newly generated offspring ones. The iterations are performed cyclically until a prefixed number of optimization. 

\subsection{\label{sec:level2}Experimental setup}
\begin{figure}[htb]
	\centering
	\includegraphics[width=0.45\textwidth]{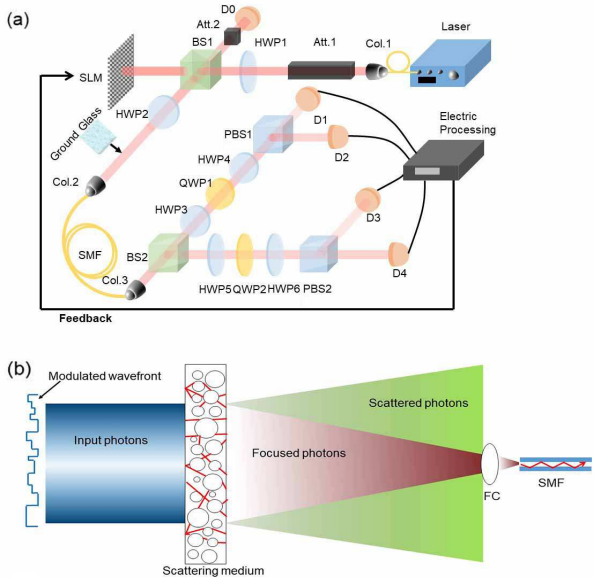}
	\caption{(a) Experimental setup for proof-of-principle QKD experiment. (b) Conceptual illustration of optimization in our scheme. Att., optical attenuator; HWP, half-wave plate; QWP, quarter-wave plate; PBS, polarization beam splitter; BS, beam splitter; SLM, spatial light modulator; Col., Fiber Collimators; SMF, Single-mode Fiber; $D0$-$D4$, single-photon detectors; FC, fiber collimator.}
	\label{fig:2setup}
\end{figure} 

We verify the validity of the method by two experiments. Firstly, we implement a proof-of-principle BB84 QKD system without scattering medium in the quantum channel (Fig. \ref{fig:2setup}(a)). Two mutually unbiased bases (MUBs), the orthogonal (Z) basis and diagonal (X) basis, are chosen as ${\ket{H}, \ket{V}}$ and ${\ket{+}, \ket{-}}$, respectively. $\ket{H}, \ket{V}, \ket{+}, \ket{-}$ corresponded to horizontal, vertical, $45^{\circ}$ and $135^{\circ}$ linear polarization, respectively. These MUBs are encoded using polarization of photons and realized by a half-wave plate (HWP2) in the system. We use a coherent laser at the wavelength of 780 nm and a 2.1 mm diameter waist. The laser pulses are attenuated to the single-photon level before transmitted into the quantum channel using an attenuator (Att.1). Another attenuator (Att.2) and a single-photon detector (SPD, $D_0$) are placed at the reflecting port of a beam splitter (BS1) to monitor the average photon number per pulse into the quantum channel. A collimator collects the single-photon signals over the scattering channels to SMF (Col.2), then the photons are incident into the decoding and detection module of Bob. The beam splitter (BS2) of Bob is used to choose measurement basis passively. The upper and down paths performed the Z basis and X basis measurements, respectively. In each path, two HWPs and a quarter-wave plate (QWP) are used to compensate for the polarization variation by the SMF. We measure the total counting rate of the single-photon detectors in Bob ($D_1$-$D_4$) and the QBER of each polarization state to characterize the system's QKD system's performance. The average detection efficiency and dark count rate of the SPD are about 55\% and 8.4 Hz, respectively.

Then, we use ground glass diffusers with 120 and 600 grits (Thorlabs, DG20-120 and DG20-600) to simulate scattering channels with different scattering strength. For the QKD system with scattering channels, we realize a single-photon level modulation of the spatial optical field to optimize the transmission efficiency. For free-space QKD systems, the single-mode-fiber (SMF) is usually used as a spatial mode filter to suppress background noise for the receiver \cite{li2019experimental,gruneisen2017modeling}. To improve the performance of the quantum channel, the coupling efficiency of the fiber receiver should also be enhanced. As is shown in Fig. \ref{fig:2setup}(b), by modulating the spatial field distribution of transmitted photons, the channel-transmission and fiber-coupling efficiencies to the single-photon signals can be optimized generation by generation using GA.

We use a spatial light modulator (SLM, Holoeye, LETO, with the precision of $0.2\pi$) to compensate the scattering effect of the quantum channel by pre-modulating the spatial phase of the photon state according to the estimating results of GA. A half-wave plate (HWP1) is adopted to rotate the incident polarization to match the SLM's polarization axis. The effective pre-modulation area of the SLM is divided into $60\times60$ blocks, each of which is $51.6\mu m\times51.6\mu m$. In the initiation phase of the algorithm, we generate 20 random patterns (population size) as the parent group, among which a blank pattern is introduced to make sure that the initial maximum coupling efficiency will not be less than that without the GA optimization. The blank pattern speeds up the optimization procedure and provides a fairer evaluation of the algorithm's optimization ability. The total single-photon detection efficiency of Bob is selected as the fitness function, which is defined as follows,
\begin{equation}
\eta\equiv\frac{D_1+D_2+D_3+D_4}{D_0}, 
\end{equation}
where $D_0-D_4$ are the single-photon counting rate of the corresponding SPDs. By choosing the total counting rate of the four SPDs, we can calculate the relative ratio of the fitness function over that of the blank pattern to evaluate the system efficiency enhancement with the optimization procedure. By dividing the transmitted photon count $D0$, we are able to mitigate intensity fluctuation of the light source and make the relative ratio coming from single-photon detection be more accurate. After optimization, we execute the QKD procedure and compare the performance with that before optimization to evaluate the effectiveness of GA procedure.

\section{\label{sec:level1}Results and Discussion}

\begin{figure}
	\centering
	\includegraphics[width=0.45\textwidth]{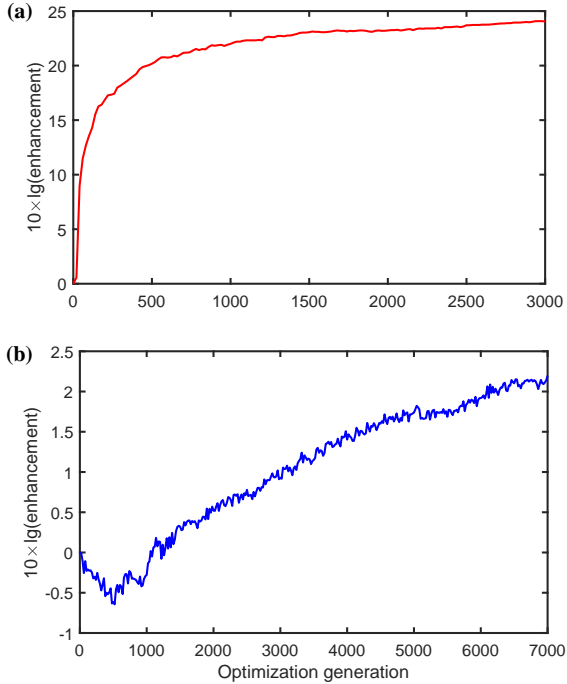}
	\caption{Optimization results of the scattering channel. (a) The enhancement of the total photon count of $D1$-$D4$ in the optimization process after a 120-grit scattering medium was added. GA algorithm has made the transmitted power coupling to the SMF a converging curve, indicating the increase of the transmission efficiency. The horizontal axis is the generation. (f) The enhancement of the total photon count of $D1-D4$ in the optimization process after a 600-grit scattering medium was added.}
	\label{fig:3GA-data}
\end{figure}

\begin{figure}
	\centering
	\includegraphics[width=0.5\textwidth]{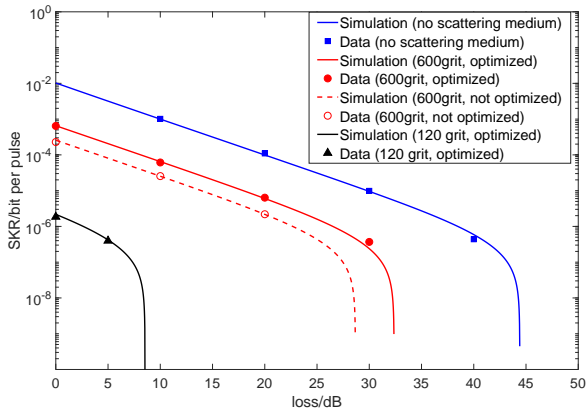}
	\caption{Secure key rates (SKRs) of the proof-of-principle QKD experiment with channel loss (where scattering loss is not included). The lines are the simulation SKRs for the quantum channels without (blue line) and with 600-grit (red solid and dashed lines) and 120-grit scattering mediums (black line), respectively. The squares, filled and hollow circles, and triangles are the corresponding experimental SKRs.
	}
	\label{fig:4SKR}
\end{figure}

\begin{figure*}
	\centering
	\includegraphics[width=\textwidth]{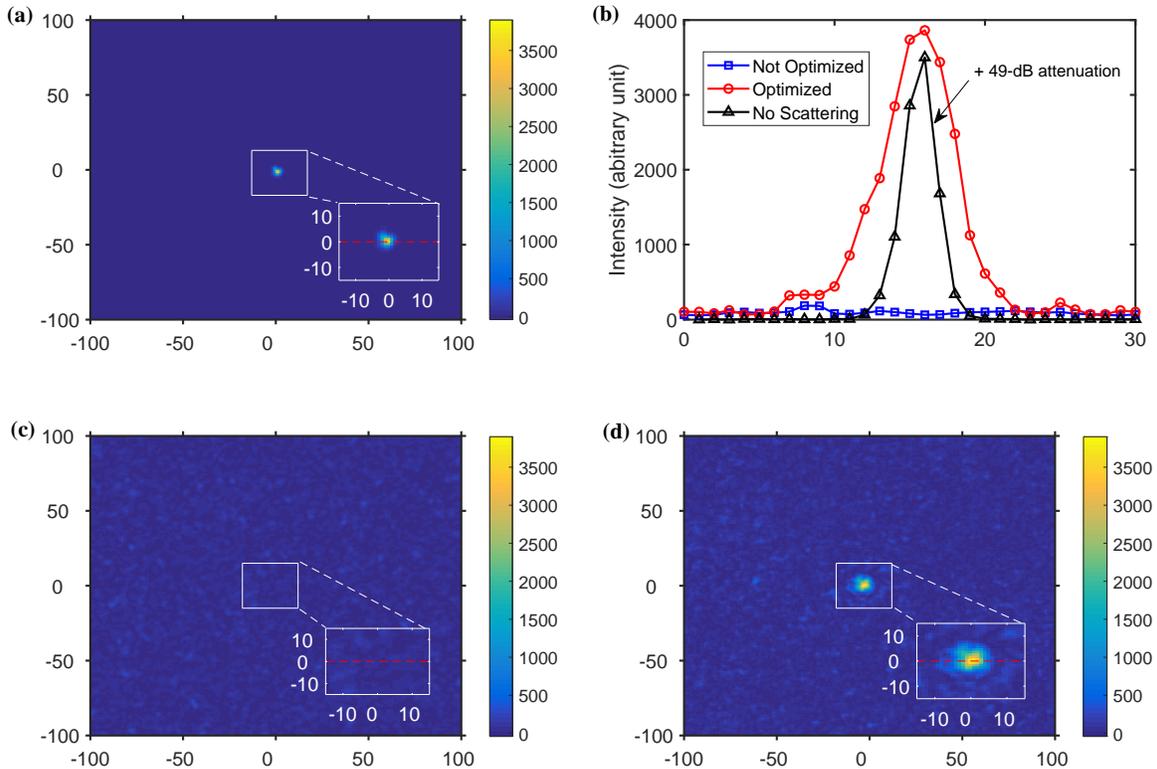}
	\caption{The beam profile of the optical fields transmitted through scattering quantum channels. (a) The beam profile without scattering medium. (b) The transverse intensity distribution of the beam without and with a scattering media of 120 grit. (c)-(d), the transmitted beam profiles through the 120-grit scattering medium without and with GA optimization, respectively. Red, blue and black lines in (b): the intensity distribution along $X$ axis (the red dashes) of the beam profiles in (a), (c) and (d), respectively. All the beam profiles ares measured with a lens (the focus length is 5cm), and when measuring without scattering medium a 49-dB attenuation is added to avoid overexposure. $x$ label in (b): pixel number. $x$ and $y$ labels in (a), (c), and (d): Number of the pixels. Pixel size: 6.4 $\mu m$. CCD exposure time in (a), (c), and (d) is 0.2 ms. Subfigures in (a), (c), and (d): The magnified intensity distribution of $30\times30$ pixels around the spot.}
	\label{fig:5Beamprofile}
\end{figure*}

When there is no scattering medium in the quantum channel, the overall transmission efficiency of the quantum channel between Alice and Bob is $64.5\%$ (about 1.9 dB), which is mainly contributed by the coupling efficiency of the fiber collimator (Col.2 in Fig. \ref{fig:2setup}(a)). After transmitting through the quantum channel with a 120-grit scattering medium, the overall transmission loss increased to 62.1 dB. However, the transmission loss over the scattering channel decreases to 38.0 dB after the optimization using the GA with 3000 generations, and the evolution process of optimization is shown in Fig. \ref{fig:3GA-data}(a). By replacing the strong scattering medium with a weaker one (600 grit), the overall transmission loss of the quantum channel becomes 16.8 dB. After optimized by GA with 7000 generations, the transmission loss decreases to 14.6 dB. The enhancement of the optimization is shown in Fig. \ref{fig:3GA-data}(b).

From the results above, we can see that the channel transmission loss under the 600-grit scattering medium without optimization is much lower than that under the 120-grit scattering medium, which indicates that the scattering effect decreases significantly. Therefore, the corresponding enhancement under a 600-grit scattering medium is much lower as well. The evolution processing show that there exists random fluctuation for adjacent generations (Figs. \ref{fig:3GA-data}(a)-\ref{fig:3GA-data}(b)), which is majorly due to the background noise. However, benefiting from the robustness of GA, this fluctuation only affects the optimization performance slightly. 

We then carry out QKD sessions after evaluating the optimization performance of the light pulses. According to Gottesman-Lo-L{\"u}tkenhaus-Preskill (GLLP) equation \cite{gottesman2004security}, the secure key generation rate of BB84 QKD is
\begin{equation}
R=max\{qQ_\mu[-f(E_\mu)H_2(E_\mu)+\Delta_1(1-H_2(e_1))],0\},
\end{equation}
where $q$ depends on the QKD protocol and is 1/2 in our experiment; $Q_\mu$ and $e_\mu$ are the gain and QBER of the signal (decoy) state; $f(e_\mu)$ is the error correction efficiency of the signal states and is set as 1.15; $H_2$ is the binary Shannon entropy, which is defined as $H_2(x)=-xlog_2(x)-(1-x)log_2(1-x)$; $\Delta_1$ and $e_1$ are the fraction and QBER of single-photon signals, respectively. We adopt the "vacuum + weak decoy-state" method \cite{wang2005beating,lo2005decoy,ma2005practical}, and the photon numbers of the signal and decoy states are set as 0.6 and 0.2, respectively. Although the decoy states have not been randomly modulated pulse by pulse in this proof-of-principle experiment, the effectiveness of this evaluation-optimization method is verified since the comparisons are under the same conditions. The simulation and experimental secure key rates are shown in Fig. \ref{fig:4SKR}, where several different loss values were chosen for each scattering channel. More details of the experimental parameters and results are shown in the appendix. 

As shown in Fig. \ref{fig:4SKR}, the introducing of the scattering media leads to a significant decrease of SKR. Especially, for strong scattering channel (120 grit), the scattered signal photons barely propagate through quantum channel and the transmission efficiency is too low to support secure key distribution (thus could not be shown in Fig. \ref{fig:4SKR}). After optimization, QKD becomes usable and the SKR is shown by the black triangles. The simulation curve (black) shows that the QKD can support a secure-key-distribution distance with an additional 8.5-dB loss. That is to say, an effective QKD channel has been built up through the optimization process for the strong scattering channel. For weak scattering channel (600 grit), the SKR enhancement is relatively small. However, it will be crucial for limit-distance communication.

In order to quantify the optimization effectiveness, we also measure the beam profiles of transmitted optical fields through the quantum channel with and without the scattering medium using a charge-coupled-devices (CCD) camera. The beam profile and intensity distribution along $x$ axis in front of Col.2 without scattering medium are shown in Fig. \ref{fig:5Beamprofile}(a) and the black triangle line of Fig. \ref{fig:5Beamprofile}(b), respectively. The intensity distribution along the $x$ axis approximates to the Gaussian distribution. After transmitting through the quantum channel with a 120-grit scattering medium, the beam profile in front of Col.2 is shown in Fig. \ref{fig:5Beamprofile}(c), which demonstrates that the optical field is seriously scattered and the intensity distribution is relatively "flat" (the blue square line in Fig. \ref{fig:5Beamprofile}(b)). After optimization with GA, the beam profile has been shaped to a sharp spot with a high contrast of about 100 to the around the area (Fig. \ref{fig:5Beamprofile}(d)). The intensity distribution of the spot is Gaussian-like (red circle line in Fig. \ref{fig:5Beamprofile}(b)), which indicates a good matching of the spatial mode to the SMF fiber and hence the transmission loss is reduced from 62.1 dB to 38.0 dB. 

The speed of proposed optimization process is mainly limited by the frame rate of the SLM. For free-space based systems under some practical channel conditions, such as some types of haze, fog or cloud, the variance of the channel characteristics are in the time scale of 1s to 1min \cite{colvero2005real,muhammad2007characterization,ijaz2013modeling}. By replacing the SLM with digital micro-mirror devices (DMD, with a frame rate of 10kHz or faster), our method can achieve an optimization from chaos within several seconds (for 1000 optimization generation). After an optimized solution has achieved, a feedback with 10-generation optimization can be fulfilled within 100 ms to track the slow changes of the scattering characteristic, which is promise to achieve a real-time correction of the practical strong scattering channel and to enhance practical free-space QKD systems.

\section{\label{sec:level1}summary}
We have applied the waveform shaping method to the QKD system through scattering quantum channels using GA based on the single-photon-level signals and optimization processing. The proof-of-principle QKD experiment clearly shows that an effective quantum link can be built through a strong scattering channel using this method, which can even make the key generation sessions from impossible to possible, which will be crucial for practical QKD under extreme field conditions. By using modulation devices with higher speed in this method, our system can be used to provide a real-time correction to the more practical strong scattering media, such as haze or fogs in the free space QKD channel. Our study offers a new perspective for expanding the application scope of practical QKD systems and will significantly enhance the global quantum communication network's deployment ability.

\section*{Funding Information}

This work has been supported by the National Key Research and Development Program of China (Grant No. 2018YFA0306400), National Natural Science Foundation of China (Grant Nos. 61627820, 61905235, 61675189, 61622506, 61822115, 61875181), Anhui Initiative in Quantum Information Technologies(Grant No. AHY030000) and the University of Science and Technology of China’s Centre for Micro and Nanoscale Research and Fabrication. K.H. also thanks CAS Pioneer Hundred Talents Program, “the Fundamental Research Funds for the Central Universities” in China, USTC Research Funds of the Double First-Class Initiative (Grant YD2030002003). F.-X.W. also thanks the support of China Postdoctoral Science Foundation (2019M652179).

\section*{Acknowledgments}

The authors thank Lei. Gong for fruitful discussion and Zhao-Di. Liu for technical support on wavefront shaping techniques. 

\section*{Disclosures}

The authors declare no conflicts of interest.

\appendix
\section{\label{sec:level1}Specific Experimental data}

We realized a proof-of-principle decoy-state BB84 QKD experiment over a strong scattering quantum channel. In the experiment, two mutually unbiased bases (MUBs), the orthogonal (Z) basis and diagonal (X) basis, were chosen as ${\ket{H}, \ket{V}}$ and ${\ket{+}, \ket{-}}$, respectively. $\ket{H}, \ket{V}, \ket{+}, \ket{-}$ corresponded to horizontal, vertical, $45^{\circ}$ and $135^{\circ}$ linear polarization, respectively. The single-photon count rate and the QBER of each polarization state were measured at Bob's SPDs to characterize the performance of the QKD system. We estimated the secure key rates by the "vacuum + weak decoy state" method \cite{wang2005beating,lo2005decoy,ma2005practical}. The 0.6 and 0.2 photon per pulse were selected as the signal and decoy states, respectively.  The lower bound of yield $Y_1$ and the upper bound of QBER $e_1$ of the single photon signals were estimated from the yield and the QBER of both signal and decoy states as the following equations.
\begin{equation}
Y_1\geq Y_1^{L,\nu,0}=\frac {\mu}{\mu\nu-\nu^2}(Q_\nu e^\nu-Q_\mu e^\mu\frac{\nu^2}{\mu^2}-\frac{\mu^2-\nu^2}{\mu^2}Y_0)
\end{equation} 

\begin{table*}[hbt]
	\flushleft
	\small
	\caption{\bf The experimental data of the QKD procedure with the decoy-state method. Labels: Scatt Loss: Scattering-induced channel loss. Total Loss: The total loss in the channel including the scattering-induced channel loss and the simulated channel loss by the attenuator. No Scatt: No scattering media. R: Secure Key Rate.}
	\begin{tabular}{ccccccccc}
		\toprule
		&Optimization &Scatt Loss(dB) &Total Loss(dB)   &$Q_\mu$	   &$Q_\nu$	  &$E_\mu(\%)$	   &$E_\nu(\%)$	             &R(per pulse) \\
		\midrule
		\multirow{3}*{120grit}  &Without	 & 62.1          	&62.1  	&$2.2\times10^{-7}$	           &$1.2\times10^{-7}$    	  &$24$	   &$37$	     &None  \\
                \cmidrule(r){2-9}
                &\multirow{2}*{With}        &38.0	     	&38.0	&$1.34\times10^{-5}$	   &$4.50\times10^{-6}$	  &$1.63$	   &$2.79$	     &$1.85\times10^{-6}$ \\
		&	                                &38.0        	 &43.0	&$4.67\times10^{-6}$          &$1.71\times10^{-6}$ 	  &$2.95$     &$6.84$	     &$3.96\times10^{-7}$ \\
		\midrule
		\multirow{8}*{600grit} 
		&\multirow{4}*{Without}	 & 16.8        	&16.8	&$1.8989\times10^{-3}$	   &$6.2211\times10^{-4}$	  &$1.72$	  &$2.51$	     &$2.29\times10^{-4}$ \\
		&	 & 16.8       	&26.8	&$1.862\times10^{-4}$      	   &$6.26\times10^{-5}$	  &$1.87$ 	   &$1.99$	     &$2.55\times10^{-5}$ \\
		&	 & 16.8          	&36.8	&$1.93\times10^{-5}$      	   &$6.49\times10^{-6}$	  &$2.27$ 	   &$3.14$	     &$2.19\times10^{-6}$  \\
		&	 & 16.8            	&46.8	&$1.99\times10^{-6}$      	   &$8.0\times10^{-7}$	  &$6.00$ 	   &$12.56$	     &None \\
                \cmidrule(r){2-9}
                &\multirow{4}*{With}	         &14.6	   	&14.6	&$3.2224\times10^{-3}$	   &$1.0862\times10^{-3}$	  &$0.80$	  &$0.77$	     &$6.43\times10^{-4}$ \\
		&	         &14.6      	 	&24.6	&$3.259\times10^{-4}$	   &$1.069\times10^{-4}$	  &$0.77$	   &$0.81$	     &$6.13\times10^{-5}$ \\
		&	         &14.6   	       &34.6		&$3.15\times10^{-5}$	   &$1.10\times10^{-5}$	  &$1.00$	   &$1.54$	     &$6.36\times10^{-6}$ \\
		&               &14.6  	      	&44.6	&$3.38\times10^{-6}$	   &$1.28\times10^{-6}$	  &$3.37$	   &$7.17$	     &$3.68\times10^{-7}$ \\
		\midrule
		\multirow{4}*{No Scatt.} &\multirow{4}*{-}               &0	       		&11.9       &$5.8597\times10^{-3}$       &$1.9355\times10^{-3}$	  &$1.02$   &$1.14$    &$1.02\times10^{-3}$ \\
		&               &0         		&21.9  	&$5.764\times10^{-4}$	   &$1.961\times10^{-4}$	  &$1.05$	   &$0.97$	     &$1.10\times10^{-4}$ \\
		&               &0          		&31.9  	&$5.87\times10^{-5}$	   &$1.94\times10^{-5}$	  &$1.13$	   &$1.44$	     &$9.86\times10^{-6}$ \\
		&                &0        		&41.9  		&$6.13\times10^{-6}$	   &$2.07\times10^{-6}$	  &$2.68$	   &$6.22$	     &$4.37\times10^{-7}$ \\
		
		\bottomrule
	\end{tabular}
	\label{tab:QKDexpe_specific}

\end{table*}

\begin{equation}
e_1\leq\frac{E_\nu Q_\nu e^\nu-e_0Y_0}{Y_1^{L,\nu,0}\nu}
\end{equation}
Then we calculated the upper bound of average single photon fraction $\Delta_1$ in the GLLP formula (Eq.3 in the main text) by 
\begin{equation}
\Delta_1\geq\frac{Y_1^{L,\nu,0}\mu e^{-\mu}}{Q_\mu},
\end{equation}
where $Q_\mu$, $E_\mu$, $Q_\nu$ and $E_\nu$ are the average yields and QBERs of the signal and decoy states. The experimental data with/without optimization and with different channel losses were measured under each scattering length. The specific experimental values of the losses, $Q_\mu$, $E_\mu$, $Q_\nu$, $E_\nu$ and the secure key rates under different scattering lengths are shown in the table below.

\bibliography{references}

\end{document}